\documentstyle[11pt,aaspp4,flushrt]{article} 
\input psfig
\begin{document}
\title{Ultra-Relativistic Blast Wave: Stability and Strong Non-Universality} 
\author{Andrei Gruzinov}
\affil{Institute for Advanced Study, School of Natural Sciences, Princeton, NJ 08540}

\begin{abstract}
Linear eigenmodes of a spherically symmetrical ultra-relativistic blast wave (the Blandford-McKee, BMK, solution) are calculated. The BMK solution is shown to be stable and strongly non-universal. It is stable because all the eigenmodes decay. Non-universality of the BMK solution (BMK is not an intermediate asymptotic) is a consequence of causality. In terms of eigenmodes -- some eigenmodes decay too slowly. For each degree $l$, there exists an eigenmode which decays at the smallest rate. The amplitude of this eigenmode, defined as relative perturbation of energy behind the shock front, is constant at early time. Later, when the blast wave slows down to $\Gamma \approx l/10$, an oscillatory decay commences, and the amplitude drops to less than 10\% of the initial value at $\Gamma \approx l/100$. The non-universality is surprisingly strong. Near the end of the ultra-relativistic stage, perturbations with $l\lesssim 200$, that is more than 10,000 different harmonics, are suppressed by less than a factor of 10. Spherical symmetry is only reached by the time the blast wave slows down to non-relativistic velocities, when the Sedov/Taylor/von Neumann solution sets in. 

\end{abstract}

\section{Ultra-relativistic blast waves, GRBs,  and their aftreglows}
It is thought that ultra-relativistic blast waves occur naturally, and GRB afterglows come from synchrotron-emitting ultra-relativistic blast waves (Piran 1999).

A spherically symmetrical ultra-relativistic blast wave (the BMK solution) was given by Blandford and McKee (1976). We will show that this solution is stable but strongly non-universal (\S 3). In \S 2 we derive the basic equations, it includes a simple derivation of the BMK solution.

Qualitatively, non-universality follows from causality. For the BMK solution, the shock front moves with Lorentz factor $\Gamma \propto t^{-3/2}$. Consider a light signal propagating along the front, starting at a polar angle $\theta =0$. Then, with $c=1$,  $(dr/dt)^2+(rd\theta /dt)^2=1$, which, for $\Gamma \gg 1$, gives $\theta ={2\over 3}\Gamma ^{-1}$. Two regions of the shock front, separated by angle $\theta$, do not talk to each other until the blast wave slows down to $\Gamma \sim \theta ^{-1}$. 

It turns out that non-universality is very strong (\S 3). The non-universality of ultra-relativistic blast waves should be observed as non-universality of early (ultra-relativistic) GRB aftreglows.

\section{Basic equations. The BMK solution.}
Relativistic hydrodynamics equations are written as $\partial _kT^{ik}=0$ (Landau \& Lifshitz 1987) . Here $x^k=(t,{\bf r})$ are Cartesian coordinates tied to the rest frame of the unshocked fluid. The energy-momentum is $T^{ik}=(4u^iu^k-g^{ik})p$ in the shocked fluid. The energy-momentum in the unshocked fluid, $\tau ^{ik}$, has only one non-zero component, and without loss of generality in the final result, we take $\tau ^{00}=1$. The shock front is at $F=0$, the boundary conditions follow from the hydrodynamics equations:
\begin{equation}
T^{ik}\partial _kF=\tau ^{ik}\partial _kF, ~~~~~~~~~~~F=0.
\end{equation}

The 4-velocity is $u^k=\gamma (1,{\bf v})$. In spherical polar coordinates, without loss of generality in the final result, ${\bf v}=v\hat{r} +u\hat{\theta }$. We use the three-dimensional form of the hydrodynamics equations:
\begin{equation}
\partial _0T^{00}+\partial _{\alpha }T^{0\alpha }=0,
\end{equation}
and 
\begin{equation}
\partial _0T^{0\alpha }+\partial _{\beta }T^{\alpha \beta }=0,
\end{equation}
which we project onto $\hat{r}$ and $\hat{\theta }$. Only first-order terms in $u$, $\partial _{\theta }p$, $\partial _{\theta }v$ are kept. We get
\begin{equation}
\partial _t[(4\gamma ^2-1)p]+\partial _r[4\gamma ^2vp]+8r^{-1}\gamma ^2vp+4\gamma ^2p(r\sin \theta)^{-1}\partial _{\theta }[\sin \theta u]=0,
\end{equation}
\begin{equation}
\partial _t[4\gamma ^2vp]+\partial _r[(4\gamma ^2v^2+1)p]+8r^{-1}\gamma ^2v^2p+4\gamma ^2vp(r\sin \theta)^{-1}\partial _{\theta } [\sin \theta u]=0,
\end{equation}
\begin{equation}
\partial _t[4\gamma ^2pu]+\partial _r[4\gamma ^2vpu]+12r^{-1}\gamma ^2vpu+r^{-1}\partial _{\theta }p=0.
\end{equation}

We parameterize the shock front as $F=r-R(t,\theta )=0$. Then eq. (1) gives, up to first order in $u$, $\partial _{\theta }R$:
\begin{equation}
4\gamma ^2p[\partial _tR-v]-p\partial_tR=\partial _tR
\end{equation}
\begin{equation}
4\gamma ^2v[\partial _tR-v]-1=0.
\end{equation}
\begin{equation}
4\gamma ^2u[\partial _tR-v]+R^{-1}\partial _{\theta }R=0.
\end{equation}

We introduce a new independent variable, $x=t-r$ ($\partial _t\rightarrow \partial _t+\partial _x$, $\partial _r\rightarrow -\partial _x$). We will assume $x\ll t,~r$. We parameterize the shock position as $R=t-x_s(t,\theta )$. Denote $\gamma ^2\equiv q$. Then, 
\begin{equation}
v=1-{1\over 2q}-{1\over 8q^2}+O(q^{-3})
\end{equation}
Keeping two leading orders in $q^{-1}$, we get
\begin{equation}
\partial _t[(4q-1)p]+\partial _x[(1+{1\over 2q})p]+8r^{-1}(q-{1\over 2})p+4qp(r\sin \theta)^{-1}\partial _{\theta } [\sin \theta u]=0,
\end{equation}
\begin{equation}
\partial _t[(4q-2)p]+\partial _x[(1-{1\over 2q})p]+8r^{-1}(q-1)p+4(q-{1\over 2})p(r\sin \theta)^{-1}\partial _{\theta } [\sin \theta u]=0,
\end{equation}
\begin{equation}
\partial _t[4qpu]+\partial _x[(2+{1\over 2q})pu]+12r^{-1}(q-{1\over 2})pu+r^{-1}\partial _{\theta }p=0.
\end{equation}

In the leading order in $q^{-1}$, these can be written as
\begin{equation}
\partial _t(qp)+2t^{-1}qp+{1\over 4}\partial _xp=-qp(t\sin \theta)^{-1}\partial _{\theta } [\sin \theta u],
\end{equation}
\begin{equation}
\partial _tp+4t^{-1}p+\partial _x(q^{-1}p)=-2p(t\sin \theta)^{-1}\partial _{\theta } [\sin \theta u],
\end{equation}
\begin{equation}
\partial _tu+t^{-1}u+{1\over 2q}\partial _xu+{u\over 4qp}\partial _xp+{1\over 4qpt}\partial _{\theta }p=0.
\end{equation}
The boundary conditions, at $x=x_s$:
\begin{equation}
q={1\over 4\partial _tx_s},~~~~~p={1\over 3\partial _tx_s},~~~~~u={\partial _{\theta }x_s\over t}.
\end{equation}

\subsection {BMK solution}
Under spherical symmetry, $u=0$, $\partial _{\theta }=0$, eqs.(14), (15), and (17) are written as
\begin{equation}
\partial _t(qp)+{2\over t}qp+{_1\over ^4}\partial _xp=0,
\end{equation}
\begin{equation}
\partial _tp+{4\over t}p+\partial _x(q^{-1}p)=0,
\end{equation}
\begin{equation}
q={1\over 4\partial _tx_s},~~~~~p={1\over 3\partial _tx_s}.
\end{equation}

These have a self-similar BMK solution 
\begin{equation}
q={\Gamma ^2\over 2\chi }, ~~~~~~~~~p={2\Gamma ^2\over 3\chi ^{17/12}},~~~~~~~~x_s={t\over 8\Gamma ^2},
\end{equation}
\begin{equation}
\chi={x\over x_s}, ~~~~~~~~~\Gamma ^2={17E\over 8\pi t^3},
\end{equation}
$E$ is the energy of the explosion.

\section{Linear eigenmodes}
We write linear perturbations of the BMK solution as 
\begin{equation}
{\delta q\over q}\equiv q(x,t)Y_{l0}(\theta ),~~~{\delta p\over p}\equiv p(x,t)Y_{l0}(\theta ),~~~
{\delta x_s\over x_{s}}\equiv \delta (t)Y_{l0}(\theta ),~~~\delta u\equiv u(x,t){dY_{l0}(\theta )\over d\theta }.
\end{equation}

We use $\chi$, and $t$ as independent variables ($\partial _t\rightarrow \partial _t-4t^{-1}\chi \partial _{\chi }$, $\partial _x\rightarrow 8\Gamma ^2t^{-1}\partial _{\chi }$). From (14)-(16), we get for the perturbations 
\begin{equation}
t\partial _tq+t\partial _tp-4\chi \partial _{\chi }q+{_{17}\over ^3}q=l(l+1)u,
\end{equation}
\begin{equation}
t\partial _tp+12\chi \partial _{\chi }p-16\chi \partial _{\chi }q+{_{20}\over ^3}q=2l(l+1)u,
\end{equation}
\begin{equation}
t\partial _tu+4\chi \partial _{\chi }u-{_{14}\over ^3}u=-{\chi \over 2\Gamma ^2}p.
\end{equation}
The boundary conditions at $\chi =1$ are 
\begin{equation}
q=-{_1\over^4}t\partial _t\delta ,~~~~~  p=-{_1\over^4}t\partial _t\delta +{_5\over ^{12}}\delta ,~~~~~u={\delta \over 8\Gamma ^2}.
\end{equation}

We perform the last change of independent variables, $t\rightarrow e^t$, $\chi \rightarrow e^{4x}$. We denote $t$ and $x$ derivatives by an over-dot and a prime. With no loss of generality in the final answer, we take such an explosion energy that $\Gamma =t^{-3/2}$. The final form of equations and boundary conditions is:

\begin{equation}
x>0:~~~\dot{q}+3q'-3p'-q=-l(l+1)u,
\end{equation}
\begin{equation}
~~~~~~~~~~~~~\dot{p}+3p'-4q'+{_{20}\over ^3}q=2l(l+1)u,
\end{equation}
\begin{equation}
~~~~~~~\dot{u}+u'-{_{14}\over ^3}u=-{_1\over^2}e^{3t+4x}p.
\end{equation}
\begin{equation}
x=0:~~~\dot{p}=\dot{q}-{_5\over^3}q, ~~~u={_3\over^{10}}e^{3t}(p-q).
\end{equation}

There are two different cases. Spherical, $l=0$, and non-spherical, $l\ne 0$. All non-spherical perturbations are isomorphic, because (28)-(31) are invariant under $l(l+1)\rightarrow \lambda l(l+1)$, $u\rightarrow u/\lambda$, $t\rightarrow t-\ln \lambda /3$.

\subsection{Spherical perturbations}
For spherical perturbations, eqs. (28)-(31) have a time independent solution, $q=0$, $p=1$, $u=0$. This is the slowest decaying eigenmode. This eigenmode is trivial -- the difference between BMK solutions of different energies. There are two other $l=0$ eigenmodes. One can check that (28)-(31) have solutions $\propto e^{-4t-4x}$ and $\propto e^{-5t-6x}$. These decay algebraically (that is algebraically in the old time variable). These eigenmodes are also trivial -- the differences between the BMK solution and BMK solutions propagating from $t\ne 0$ and $r\ne 0$. Numerical simulations (\S 3.2) show that BMK solution is the intermediate asymptotic of a spherical explosion.

\subsection{Numerical simulations}
Numerical simulations of eqs. (28)-(31) were performed after introducing ``Riemann invariants'' (Landau \& Lifshitz 1987). Let 
\begin{equation}
f=q+{_{\sqrt{3}}\over ^2}p, ~~~~~~~~g=q-{_{\sqrt{3}}\over ^2}p.
\end{equation}
Then
\begin{equation}
\dot{f} =(2\sqrt{3} -3)f'+({_1\over ^2}-{_5\over ^{\sqrt{3}}})(f+g)+(\sqrt{3} -1)l(l+1)u,
\end{equation}
\begin{equation}
\dot{g} =-(2\sqrt{3} +3)g'+({_1\over ^2}+{_5\over ^{\sqrt{3}}})(f+g)-(\sqrt{3} +1)l(l+1)u,
\end{equation}
\begin{equation}
\dot{u} =-u'+{_{14}\over ^3}u-{_1\over ^{2\sqrt{3}}}e^{3t+4x}(f-g).
\end{equation}

Since $g$ and $u$ propagate to larger $x$, one needs two boundary conditions at $x=0$; these are given by eq. (31). We need one boundary condition at the end of the simulation domain $x=x_{\rm max}$. We checked that this boundary condition is irrelevant if $x_{\rm max}$ is large enough.

Results, the time evolution of $p(x=0,t)$, are shown in fig. 1.

\begin{figure}[htb]
\psfig{figure=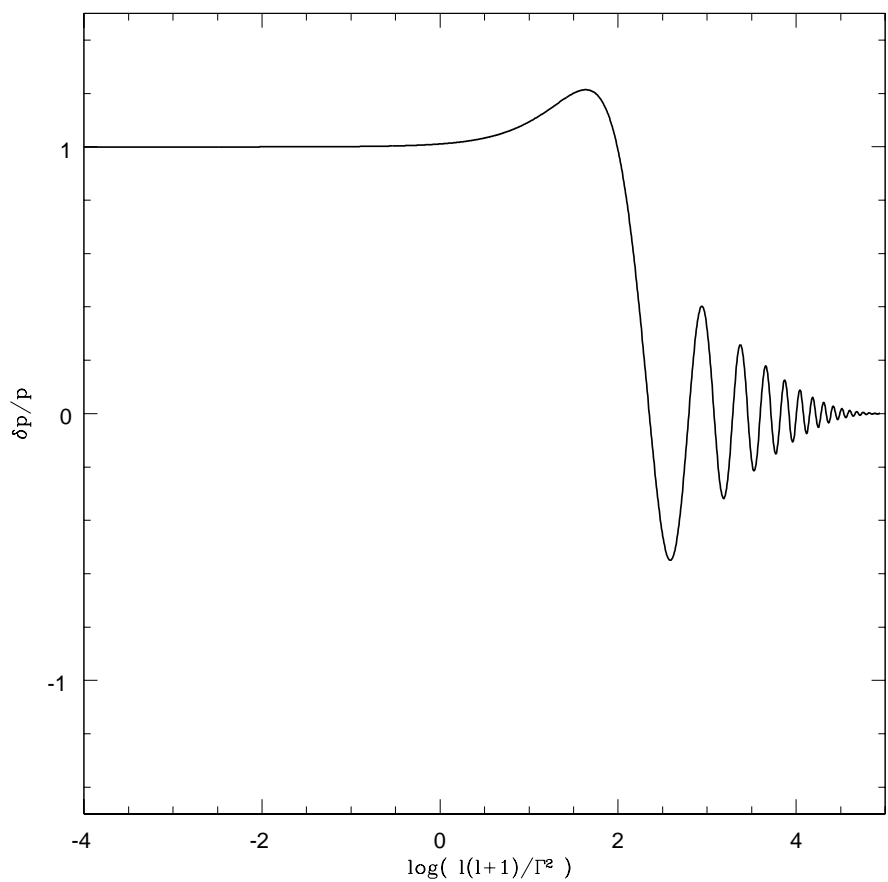,width=5in}
\caption{Relative pressure perturbation (normalized to relative pressure perturbation at $t=0$) as a function of $l(l+1)/\Gamma ^2$.}
\end{figure}

\acknowledgements This work was supported by the Keck Foundation and NSF PHY-9513835.

\end{document}